\documentclass[aps,prb,longbibliography,superscriptaddress,twocolumn,showkeys,amsmath,amssymb,floatfix]{revtex4-2}

\usepackage[english]{babel}
\usepackage{amsmath,amssymb,bbm,mathrsfs,bm,braket,color,graphicx,comment,amsfonts,dsfont}
\usepackage{siunitx,physics}
\usepackage[colorlinks,citecolor=blue,urlcolor=blue]{hyperref}

\begin{document}

\title{Horizon physics of quasi-one-dimensional tilted Weyl cones on a lattice}
\author{Viktor K\"{o}nye}
\affiliation{Institute for Theoretical Solid State Physics, IFW Dresden and W\"{u}rzburg-Dresden Cluster of Excellence ct.qmat, Helmholtzstr. 20, 01069 Dresden, Germany}
\author{Corentin Morice}
\affiliation{Institute for Theoretical Physics and Delta Institute for Theoretical Physics, University of Amsterdam, 1090 GL Amsterdam, The Netherlands}
\affiliation{Laboratoire de Physique des Solides, CNRS UMR 8502, Université Paris-Saclay, F-91405
Orsay Cedex, France}
\author{Dmitry Chernyavsky}
\affiliation{Institute for Theoretical Solid State Physics, IFW Dresden and W\"{u}rzburg-Dresden Cluster of Excellence ct.qmat, Helmholtzstr. 20, 01069 Dresden, Germany}
\author{Ali G. Moghaddam}
\affiliation{Department of Physics, Institute for Advanced Studies in Basic Sciences (IASBS), Zanjan 45137-66731, Iran}
\affiliation{Computational Physics Laboratory, Physics Unit, Faculty of Engineering and
Natural Sciences, Tampere University, FI-33014 Tampere, Finland}
\author{Jeroen van den Brink}
\affiliation{Institute for Theoretical Solid State Physics, IFW Dresden and W\"{u}rzburg-Dresden Cluster of Excellence ct.qmat, Helmholtzstr. 20, 01069 Dresden, Germany}
\affiliation{Institute for Theoretical Physics, TU Dresden, 01069 Dresden, Germany
}
\author{Jasper van Wezel}

\affiliation{Institute for Theoretical Physics and Delta Institute for Theoretical Physics, University of Amsterdam, 1090 GL Amsterdam, The Netherlands}
\date{\today}

\begin{abstract}
    To simulate the dynamics of massless Dirac fermions in curved spacetimes with one, two, and three spatial dimensions we construct tight-binding Hamiltonians with spatially varying hoppings. These models represent tilted Weyl semimetals where the tilting varies with position, in a manner similar to the light cones near the horizon of a black hole. We illustrate the gravitational analogies in these models by numerically evaluating the propagation of wave packets on the lattice and then comparing them to the geodesics of the corresponding curved spacetime. We also show that the motion of electrons in these spatially varying systems can be understood through the conservation of energy and the quasi-conservation of quasimomentum. This picture is confirmed by calculations of the scattering matrix, which indicate an exponential suppression of any noncontinuous change in the quasimomentum. Finally, we show that horizons in the lattice models can be constructed also at finite energies using specially designed tilting profiles.
\end{abstract}

\maketitle

\section{Introduction}
Analogies in physics, between seemingly unrelated systems, can not only lead to advances in the understanding of these systems, but also contribute to the development of new applications.
Among the known analogies, a particularly fruitful connection has been made between gravitational physics and condensed matter systems.
Such an analogy was first proposed to study the extraordinary consequences of the curvature of spacetime on quantum fields, namely the Hawking and Unruh radiations, which are in practice too weak to be measured \cite{unruh1981, Crispino2008, hawking1974nature, hawking1975, Page2005}.
This has attracted considerable amount of interest in recent years from various directions and gravitational analogies have been considered in many contexts including electronic, acoustic, optical and even magnetic and superconducting settings \cite{Barcelo2001, philbin2008, Carusotto2008, barcelo2011analogue, Boada2011, Riera2012, sindoni2012emergent, nguyen2015, Minar2015, Celi2017, Calabrese2017, duine2017, Rodriguez2017, Kosior2018, kollar2019hyperbolic, Nissinen2020, Boettcher2020, lapierre2020, Jafari2019, boettcher2021Hyperbolic, Mula2021, Sabsovich2021, de2021artificial, Stalhammar2021}.
Particularly, implementations using Bose-Einstein condensates have been used to mimic Hawking and Unruh radiations \cite{Weinfurtner2011, steinhauer2016, Hu2019}.

Recent attempts have been made to exploit developments in the prediction and synthesis of Weyl semimetals (WSMs), whose low-energy electronic states can be described by the Weyl equation \cite{Armitage2018}.
In particular, it has been shown that a position-dependent tilting of the Weyl cone in a way to create neighboring regions with type-I and type-II WSM can lead to a black hole analogue \cite{Volovik:2016kid, volovik2003universe, Weststrom2017, guan2017artificial, huang2018black, Ojanen2019}.
This analogy can be understood by thinking of the Weyl cones in the material as the light cones of a curved spacetime, thus the boundary of the WSMs of different types as the horizon. Exactly at the transition point between type-I and type-II, the Weyl cones are tangent to the zero-energy surface, a situation called type-III WSM. One study has shown that the band structure of Zn$_2$In$_2$S$_5$ has such a property that makes it promising for the realization of a black hole horizon \cite{huang2018black}.
Proposals for tilting the cone as a function of real space include the use of structural distortions, spin textures, and external position-dependent driving \cite{Vozmediano2018,Jeroen2019rotation,Bardarson2019,Weststrom2017,long2020,lee2016}. Most works until recently assumed that, given that the tilting changes smoothly, one can define band dispersions varying as a function of space.
Since WSMs are defined on a lattice, and the tilting variation implies the lack of translational symmetry, and therefore of a reciprocal space, this kind of assumption is imprecise. This was addressed subsequently, by studying a class of tight-binding models with position-dependent nearest-neighbor hopping concentrating on the single-band one-dimensional (1D) cases \cite{morice2021prr, moghaddam2021engineering}.

Here we go beyond previous works by introducing two-band tilted Weyl cones defined on lattices, in one, two and three spatial dimensions and investigate their properties.
Taking into account the lattice explicitly is motivated by physical Weyl semimetals having band structures that are ultimately defined on discrete lattices, which play an important role near horizons when wave-packets start slowing down exponentially~\cite{morice2021prr}. Moreover, lattice formulations allow for numerically tracking wave-packet dynamics with the very efficient Chebyshev expansion method \cite{Kosloff1994,Weisse2006}. 
We discuss the consequences of different ways of creating Weyl cones on a lattice as such models are not unique.
We connect the low momentum limit of these models to gravitational physics and show they are equivalent to Dirac equations in a curved spacetime background.
We emphasize that, due to the two-band nature of our models, the connection to the relativistic Dirac equation is more explicit than previously studied incarnations. Then, we explore the propagation of wave packets in these models and compare them to geodesics of gravitational systems.
Finally, we calculate the scattering matrix in the 1D models and discuss the large quantitative and qualitative differences arising between the different types of models.

\section{Tilted Weyl Cone Models}

In order to simulate horizons in lattice models we construct inhomogeneous tight-binding models hosting tilted Weyl cones, whose tilting depends on the position in real space. The horizon appears at the points where the tilted cone goes through the Fermi level, i.e.\ where the WSM goes from type-I to type-II.  We want the local low energy effective Hamiltonian to correspond to the following tilted Weyl continuum Hamiltonian
\begin{equation}
        \label{eq:tiltedWeylH}
        H=\vb*{\sigma}\cdot \vb{k} - \sigma_0(\vb{t}\cdot\vb{k}),
\end{equation}
\begin{figure}
\centering
\includegraphics[width=8.6cm]{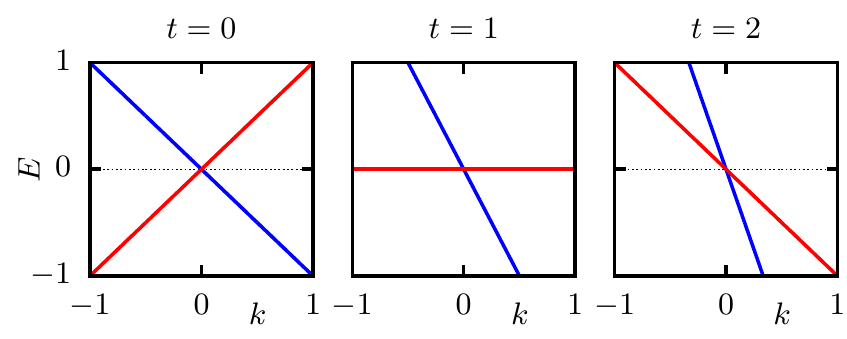}
\caption{\label{fig:tilted_Weyl} Energy dispersion of a tilted Weyl Hamiltonian in 1D as in Eq. \eqref{eq:tilteddisp} for three different tilting parameters $t$.}
\end{figure}
where $\vb{k}$ is the quasi-momentum, $\sigma_i$ are the Pauli matrices acting on a pseudospin representing different orbitals, and $\vb{t}$ is the tilting vector. The energy dispersion of this Hamiltonian is
\begin{equation}
        \label{eq:tilteddisp}
        E_\pm = \pm|\vb{k}|-\vb{t}\cdot\vb{k}.
\end{equation}
For $\vb{t}=\vb{0}$ we get an untilted Weyl cone. In the one-dimensional case (see Fig. \ref{fig:tilted_Weyl}), for $t>0$ the spectrum is tilted clockwise and at $t=1$ one of the branches of the cone becomes completely flat and the group velocity in this branch becomes zero.

In this section, first we discuss how to construct simple lattice models in one- (1D), two- (2D) and three-dimensions (3D) that host one or several tilted Weyl cones. Then, we show how these Hamiltonians are connected to the Dirac equation in a curved space-time when the tilting changes as a function of position. With this we get lattice models that describe the motion of electrons in a curved background.

\subsection{1D lattice models}

We start with presenting 1D models that have tilted Weyl cones at low energies. In principle, there are many ways to put the Eq. \eqref{eq:tiltedWeylH} Hamiltonian on a lattice with the same low energy effective Hamiltonian around $k=0$, but with different behavior at $|k|\gg0$. Here, we study the following two models
\begin{subequations}
\label{eq:realham}
\begin{align}
\label{eq:realham1}
        \mathcal{H}_1^{\text{1D}} &= \sum\limits_{x} \frac{i}{2}c^\dag_{x+1}(\sigma_x - t\sigma_0) c^{\vphantom{\dag}}_x + \mathrm{h.c.} ,\\
        \mathcal{H}_2^{\text{1D}} &= \sum\limits_{x} c^\dag_{x}\frac{\sigma_z}{2} c^{\vphantom{\dag}}_x + c^\dag_{x+1}\left[-\frac{\sigma_z}{2}+\frac{i}{2}(\sigma_x - t\sigma_0)\right] c^{\vphantom{\dag}}_x + \mathrm{h.c.},
\end{align}
\end{subequations}
where $c_x=(c_{x\uparrow},c_{x\downarrow})$ is the annihilation operator of an electron at site $x\in\mathbb{Z}$ with pseudo-spin $\uparrow$ or $\downarrow$.

\begin{figure}
\centering
\includegraphics[width=8.6cm]{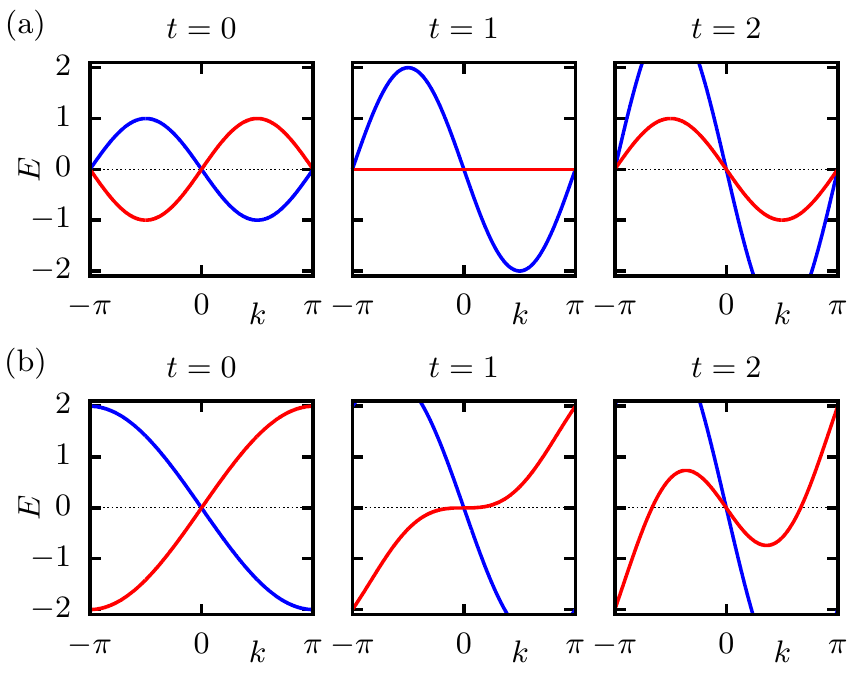}
\caption{\label{fig:disprel_1d} Energy dispersion of (a) $H_1^{\text{1D}}$ as in Eq. (\ref{eq:E1}) and (b) $H_2^{\text{1D}}$ as in Eq. (\ref{eq:E2}) for three different tilting parameters $t$. The three tiltings correspond to the untilted type I node, the node at the horizon and the overtilted type II node.}
\end{figure}

For a constant position-independent tilt, the Bloch Hamiltonians of these after Fourier transformation are given as
\begin{subequations}
 \label{eq:H1H2}
\begin{align}
        H_1^{\text{1D}} &= (\sigma_x - t\sigma_0)\sin{k},\\
        H_2^{\text{1D}} &=(\sigma_x - t\sigma_0)\sin{k} + \sigma_z (1-\cos{k}).
\end{align}
\end{subequations}
which give the following dispersion relations (see Fig.~\ref{fig:disprel_1d})
\begin{subequations}
\begin{align}
        \label{eq:E1}
        E_{1\pm}^{\text{1D}} &= \pm |\sin{k}| - t\sin{k},\\
        \label{eq:E2}
        E_{2\pm}^{\text{1D}} &= \pm \sqrt{\sin^2{k}+(1-\cos{k})^2} - t\sin{k}.
\end{align}
\end{subequations}
Both these models converge to Eq. (\ref{eq:tiltedWeylH}) for $k\to0$ and describe the same kind of tilted Weyl cone, but they differ away from $k=0$. In particular, $H_1^{\text{1D}}$ has a second Weyl cone at the edge of the Brillouin zone, unlike $H_2^{\text{1D}}$ which only has one cone. In both cases if the tilting parameter is less (more) than one we get type I (type II) Weyl nodes. In the overtilted type II case there is an electron (hole) pocket forming to the right (left) of the node.

The main difference between the two models, that will be relevant in later results, is that the red band in the overtilted region does not cross zero between $k=0$ and $k=\pi$ in $H_1^{\text{1D}}$, but it does in $H_2^{\text{1D}}$. In the first model the electron and hole pockets extend throughout the whole Brillouin zone and they connect the two Weyl nodes, while in $H_2^{\text{1D}}$ the pockets are finite and located next to the single node, which is the situation encountered in the band structure of type II Weyl semimetals.

\subsection{2D and 3D lattice models}

\begin{figure*}
\centering
\includegraphics[width=16.44cm]{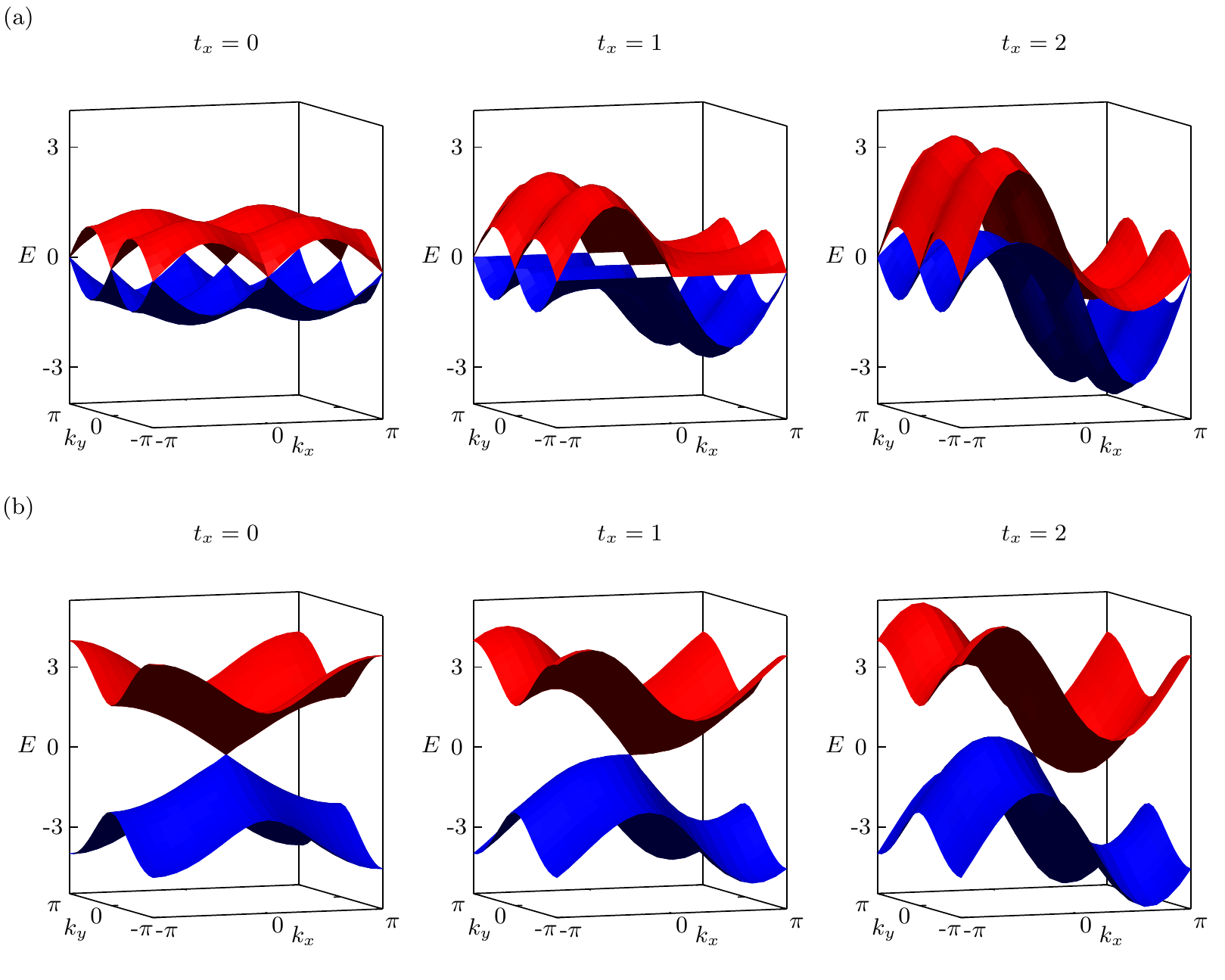}
\caption{\label{fig:disprel_2d} Energy dispersion of (a) $H_1^{\text{2D}}$ in Eq. (\ref{eq:H12D}) and (b) $H_2^{\text{2D}}$ in Eq. (\ref{eq:H22D}) for three different tilting parameters in the $x$ direction $t_x$, and $t_y=0$. The three tiltings correspond to the untilted type I node, the node at the horizon and the overtilted type II node.}
\end{figure*}

Similar to the 1D case there are infinitely many ways to create 2D and 3D lattice models with a tilted Weyl cone around $k=0$. Generalizing the two systems studied in 1D we get the following Bloch Hamiltonians in 2D
\begin{subequations}
\label{eq:H2D}
\begin{align}
        \label{eq:H12D}
        H_1^{\text{2D}} &= (\sigma_x - t_x\sigma_0)\sin{k_x}+(\sigma_y - t_y\sigma_0)\sin{k_y},\\
        \label{eq:H22D}
        H_2^{\text{2D}} &=H_1^{\text{2D}} + \sigma_z (2-\cos{k_x}-\cos{k_y}).
\end{align}
\end{subequations}
Their dispersion relations are shown in Fig. \ref{fig:disprel_2d}. $H_1^{\text{2D}}$ has 4 Weyl nodes in the entire Brillouin zone, while $H_2^{\text{2D}}$ again only has a single Weyl node.

In 3D the Hamiltonian corresponding to $H_2^{\text{1D}}$ does not exist, because the $\sigma_z$ Pauli matrix was already used. This is consistent with the Nielsen-Ninomiya theorem \cite{Nielsen1981} which does not allow a single Weyl node in a 3D lattice. The 3D equivalent of $H_1^{\text{1D}}$ is
\begin{align}
        \label{eq:H13D}
        H_1^{\text{3D}} &= \sum\limits_i (\sigma_i - t_i\sigma_0)\sin{k_i},
\end{align}
which has 8 Weyl nodes. At $k_z=0$ the dispersion relation is identical to that of $H_1^{\text{2D}}$, represented in Fig.~\ref{fig:disprel_2d}(a).

\subsection{Dirac equation in curved spacetime}
In the long-wavelength limit, the dynamics of the lattice models can be described by a low-energy continuum model. In particular, the effective Hamiltonians for the lattice Hamiltonians in Eqs. \eqref{eq:H1H2}, \eqref{eq:H2D} and \eqref{eq:H13D} can be written (using Einstein notation) as  (see Appendix \ref{app:discrete} for details)
\begin{equation}
    \label{eq:hamiltonain_le}
    H_{\text{eff}} = -i(\sigma_i-t_i\sigma_0)\partial_i-\frac{\sigma_0}{2i} \partial_i t_i,
\end{equation}
where $i$ runs on the spatial dimensions of the system ($d=1,2,3$) and $t_i(x)$ are position-dependent tilting functions in different directions.
The last term in Eq.~\eqref{eq:hamiltonain_le} is required to make the Hamiltonian Hermitian. This equation can be thought of as a Dirac equation in a curved background. To explore this gravitational analogy in a more precise way, we introduce the metric proposed in Ref.~\cite{kedem2020black}
\begin{equation}\label{metric}
    ds^2=\left(t^2 -1 \right)d\tau^2 -2 t_i dx_i d\tau +dx^2,
\end{equation}
 where $\tau$ denotes the temporal coordinate, $t^2=t_i t_i$ and $dx^2=dx_idx_i$. The massless Dirac equation for this background metric can we written in the form \cite{parker2009book}
\begin{equation}\label{eq:Dirac}
    \partial_\tau\Psi = \left(\gamma^{0 i}\partial_i-t_i\partial_i-\frac12 \partial_i t_i-\frac14 (\partial_i t_j) \gamma^{ij}\right)\Psi,
\end{equation}  
where $\gamma^0$ and $\gamma^i$ are the gamma matrices in flat space-time and by definition $\gamma^{ab}= [\gamma^a,\gamma^b]/2$ for $a,b\in (0,1,\cdots,d)$. 
To link this equation to the Hamiltonian \eqref{eq:hamiltonain_le}, we take the following representations for the gamma matrices:
\begin{eqnarray}
        \label{eq:gm1D}
        &&
        \text{in (1+1)D:} \qquad  \gamma^0=i\sigma_z, \qquad \gamma^1=\sigma_y,
        \\
        &&
        \text{in (1+2)D:} \qquad \gamma^0=i\sigma_z, \qquad \gamma^1=\sigma_y, \qquad \gamma^2=-\sigma_x,
        \nonumber
\end{eqnarray}
while in (1+3)D one may choose the Weyl representation
\begin{equation}
    \gamma^0 = 
    \begin{pmatrix}
        0 & \sigma_0\\
        -\sigma_0 & 0
    \end{pmatrix},
    \qquad \gamma^i =
    \begin{pmatrix}
        0 & \sigma_i\\
        \sigma_i & 0
    \end{pmatrix}.
\end{equation}  
The Dirac equation on the gravitational background \eqref{metric} coincides with the low-energy dynamical equation for the lattice models provided the last term in Eq.~\eqref{eq:Dirac} vanishes. In $(1+1)$D, this is always the case. In higher dimensions, we constrain the vector $t_i(x)$ to have a vanishing curl 
\begin{equation}
    \label{eq:curl}
    \partial_{i} t_{j} - \partial_{j} t_{i}=0,
\end{equation}
which cancels the last term in \eqref{eq:Dirac}. Keeping this term would produce a spatially-varying on-site potential in the tight-binding model. Since the term is proportional to the derivative of the tilting, its effect is negligible in systems with slowly varying tilting.
 
The region in space described by $t^2-1=0$ can be thought of as an event horizon. 
Notice that Eq.~\eqref{metric} represents a wide class of metrics, some of which are of particular importance in the context of gravitational physics (see, e.g., a recent discussion in \cite{volovik2021macroscopic}). 

To simulate different metrics we introduce spatial inhomogeneity in the tilting parameter. In the real space Hamiltonians \eqref{eq:realham} we make the tilting parameter position dependent. This is different from previous models where the position dependence is in the Fermi velocity \cite{Peres_2009,morice2021prr, moghaddam2021engineering}. In this paper we will only consider effectively 1D horizons, so $t_x$ will be a function of $x$ and $t_{y/z}=0$. In all cases the horizon will be defined by the points where $t_x=1$.

\section{Wave packet dynamics}

Now that we defined the systems of interest, we move on to study the propagation of wave packets in lattice models with horizons. We study the different behaviors for our models with different horizons. We will show how the world lines of wave packets in our models match the geodesics of specific metrics, and we will simulate the propagation of Gaussian wave packets based on the Chebyshev expansion method. The details of this method are explained in Appendix \ref{app:kpm}. The videos for all simulations discussed in the paper are included in the Supplemental Material (SM).

\subsection{Linear horizon in 1D}
\label{sec:1dlin}

First, we discuss 1D models with linearly changing tiltings $t_x=2x/L$, where $L$ is the length of the system. With this we linearly increase the tilting from $t=0$ to $t=2$ with the $t=1$ horizon exactly at $L/2$. The systems we study are $L=1000$ long. The starting wave packet is localized at $x_0=100$ in the red band of Fig.~\ref{fig:disprel_1d}, and propagates to the right towards the horizon. 

\begin{figure}
\centering
\includegraphics[width=8.6cm]{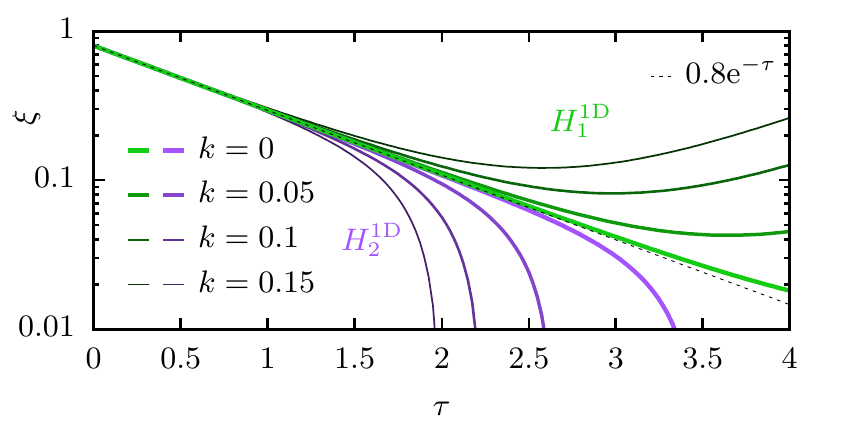}
\caption{\label{fig:1d_x_t} Propagation of wave packets in a 1D lattice with a horizon. The figure shows the center of mass of the wave packet as a function of time for different starting momenta in the systems $H_1^{\text{1D}}$ and $H_2^{\text{1D}}$ in Eq.~\eqref{eq:H1H2}. The rescaled position and time are defined as $\xi=1-2x/L$ and $\tau=2t/L$. The dashed line shows the geodesic calculated from Eq.~\eqref{eq:geodesics}. The length of the system is $L=1000$.}
\end{figure}

The red band becomes more and more flat approaching the horizon in both $H_1^{\text{1D}}$ and $H_2^{\text{1D}}$. As a result, the group velocity of the wave packet gets smaller and smaller. While the wave packet is slowing down it becomes narrower. The time dependence of the position of the wave packet is shown in Fig.~\ref{fig:1d_x_t}. For the figures we rescaled the position and time using $\xi=1-2x/L$ and $\tau=2t/L$. With this the wave packet moves from the initial position $\xi_0=0.8$ to the horizon $\xi=0$.

The wave packet dynamics results can be understood with a philosophy similar to that of the WKB approximation. Since the tilting parameter varies continuously and slowly with position we can think of the system as being locally the infinite $H_{1/2}^{\text{1D}}(\vb{k})$. Since the tilting is proportional to $\sigma_0$ the eigenvectors at different tiltings are all the same, from which it follows that states in the red or blue bands will stay eigenstates at whatever point of the chain. This means, that a wave packet that is initially in the red (blue) band will stay in the red (blue) band. The system is not strictly homogeneous, thus the momentum is not a conserved quantity, but since the inhomogeneity comes from a slow change in space, large jumps of the momentum are exponentially suppressed (this will be further clarified in Sec.~\ref{sec:S}). The quantity that is strictly conserved is the energy of the wave packet. Using the above mentioned rules together with the dispersion relations in Fig.~\ref{fig:disprel_1d}, the results of the 1D simulations can be explained.

\begin{figure}
\centering
\includegraphics[width=8.6cm]{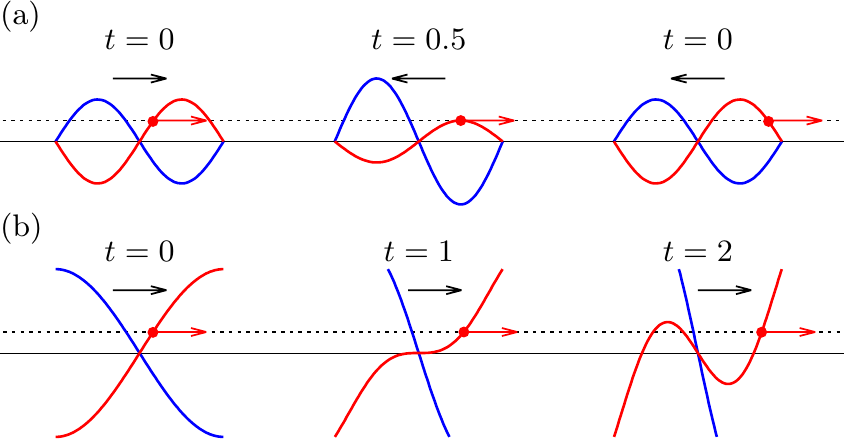}
\caption{\label{fig:1d_wkb} Schematic representation for the propagation of wave packets with $E=0.5$ in systems (a) $H_1^{\text{1D}}$ and (b) $H_2^{\text{1D}}$. The local dispersion relation is shown at three points along the trajectory of the wave packets. The red dot represents the momentum of the wave packet and the red arrow indicates the direction of change in this momentum. The black arrow shows the direction of the group velocity of the wave packet.}
\end{figure}

First, we discuss the $H_1^{\text{1D}}$ system. Focusing on the red band we see that it becomes completely flat at the horizon. If we take a packet with a small energy $E>0$, the horizon is invisible for it since after a certain tilting there is no point in the red band that is at this energy. With a continuous slow change of the momentum, the state with energy $E$ will go from $k=0$ to $k=\pi$ and the group velocity will go from positive to negative while the wave packet bounces back from the horizon (see Fig.~\ref{fig:1d_wkb}(a)). The closer the energy is to $0$, the closer the wave packet reaches the horizon and in the limiting case of $E=0$ the wave packet gets pinned to the horizon. On longer time scales because of the lattice length-scale that unavoidably comes into play~\cite{morice2021prr}, the wave packet slowly disintegrates and the center of mass will go towards the left. The behavior of the wave packets at different energies can be seen in Fig.~\ref{fig:1d_x_t} with the green curves.

The $H_2^{\text{1D}}$ system is qualitatively very different from the $H_1^{\text{1D}}$ system. Here, at every position there is an $E>0$ state in the red band. This means that a finite energy wave packet will go through the horizon with a continuous change in momentum (see Fig.~\ref{fig:1d_wkb}(b)). The zero energy wave packet will approach slowly the horizon because of the zero group velocity at the horizon but eventually it will go through it. The behavior of the wave packets at different energies can be seen in Fig.~\ref{fig:1d_x_t} with the purple curves.

The time dependence of the wave packet approaching the horizon is consistent with the results in Ref.~\cite{Morice2021}. There, a one-band model with nearest-neighbour hopping was studied, where the hopping varies with the position. This can be understood, by coarse-graining, as a model with a linear dispersion close to zero energy, whose slope varies with position. The slope was set to zero at the origin of the lattice, and finite away from it, thus mimicking the evolution of a light cone when moving away from a horizon. Since it has only one band, the hopping is exactly zero where the horizon is meant to be, meaning that the right and left side of the lattice are completely disconnected. In our two-band model this is not the case and, as we saw in the $H_2^{\text{1D}}$ system, the wave packet can propagate through the horizon. 

It was shown in Ref.~\cite{Morice2021}, both using numerical calculations and an analytical derivation based on a semiclassical approximation, that zero-energy wave packets propagating in the one-band model precisely follow the geodesics of a (1+1)D dilaton gravity. In the case where the hopping evolves linearly with position, these geodesics can be expressed as
\begin{equation}
    1-\alpha x = \mathrm{e}^{-\alpha(t-t_0)},
\end{equation}
where in our case $\alpha=2/L$. Using the previously defined scaled position and time this simply reads as
\begin{equation}
\label{eq:geodesics}
  \xi = \xi_0\mathrm{e}^{-\tau}.
\end{equation}

As we can see in Fig.~\ref{fig:1d_x_t} this dependence is present for our models too, with deviations only at larger timescales due to finite size effects, when the size of the wave packet becomes comparable to the lattice constant.

\subsection{Linear horizon in 2D and 3D}
\label{sec:2d3dlin}

\begin{figure}
\centering
\includegraphics[width=8.6cm]{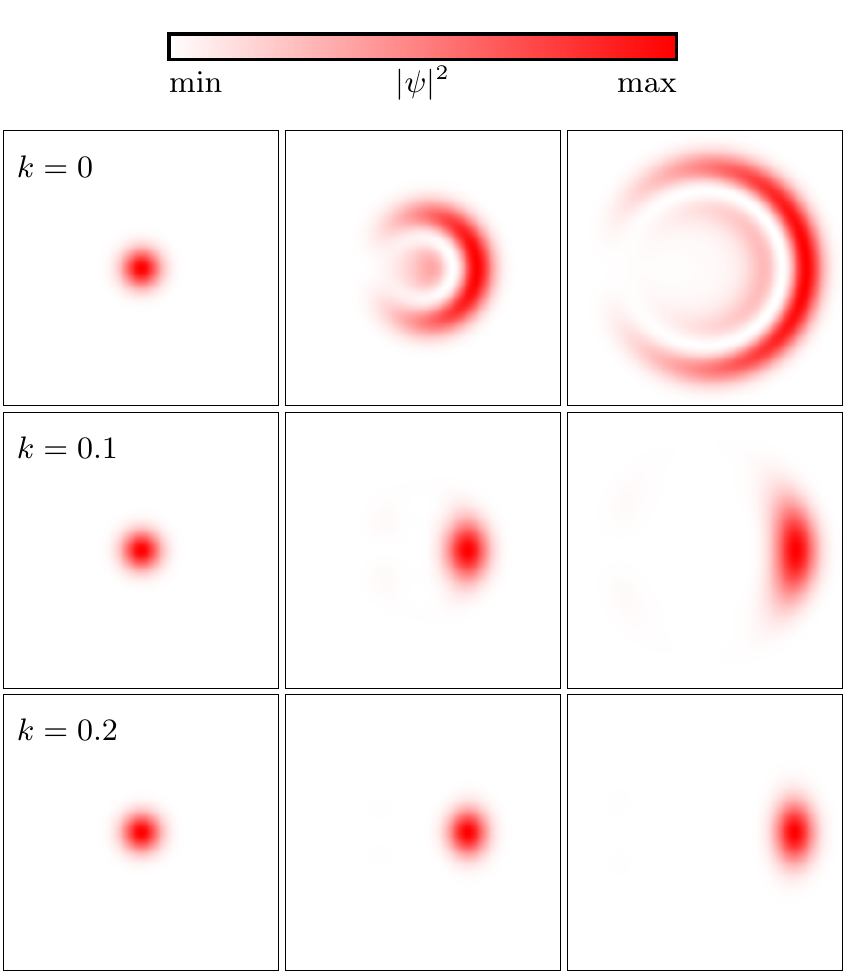}
\caption{\label{fig:2d_snapshots} Propagation of wave packets in the 2D systems without tilt. The figure shows snapshots of the wave packet for different starting momenta in the system $H_1^{\text{2D}}$. The length and width of the system are $L=500$ and $W=500$, the initial wave packet has a width of 40 sites. The anisotropy in the top row is due to the specific pseudospin configuration of the initial wave packet.}
\end{figure}

In 2D and 3D, wave packets propagate qualitatively differently from the 1D case. To understand the difference, we first consider a 2D system where $t_x=0$ and $t_y=0$ everywhere. A Gaussian wave packet centered at a large enough $k$ will propagate based on the group velocity at $k$ and will maintain its Gaussian shape. But getting closer to $k=0$ this is no longer true. A wave packet with finite size in real space will also have a finite size in momentum space. This means that at low enough momenta the wave packet encapsulates the Weyl node and thus it will have components propagating in all directions radially. This behavior of the wave packet propagation is shown in Fig.~\ref{fig:2d_snapshots} with snapshots (in the SM the full animations are available).

In all simulations we choose the pseudospin components corresponding to the eigenvector of the infinite Hamiltonian with the average momentum of the wave packet. Because the eigenvectors are momentum dependent and because the wave packet includes multiple momenta a zero momentum wave packet includes a superposition of positive and negative energy states which leads to Zitterbewegung of the electrons \cite{Schrodinger1930,Barut1981,Demikhovskii2010,Demikhovskii2008}.

\begin{figure}
\centering
\includegraphics[width=8.6cm]{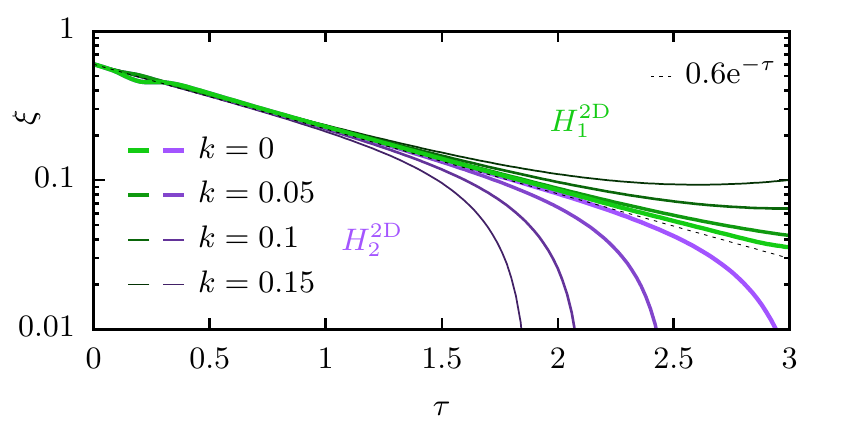}
\caption{\label{fig:2d_x_t} Propagation of wave packets in a 2D lattice with a horizon. The figure shows the center of mass of the wave packet along the $x$ axis as a function of time for different starting momenta in the systems $H_1^{\text{2D}}$ and $H_2^{\text{2D}}$ in Eq.~\eqref{eq:H2D}. The rescaled position and time are defined as $\xi=1-2x/L$ and $\tau=2t/L$. The dashed line shows the geodesic calculated from Eq.~\eqref{eq:geodesics}. The length and width of the system are $L=500$ and $W=500$.}
\includegraphics[width=8.6cm]{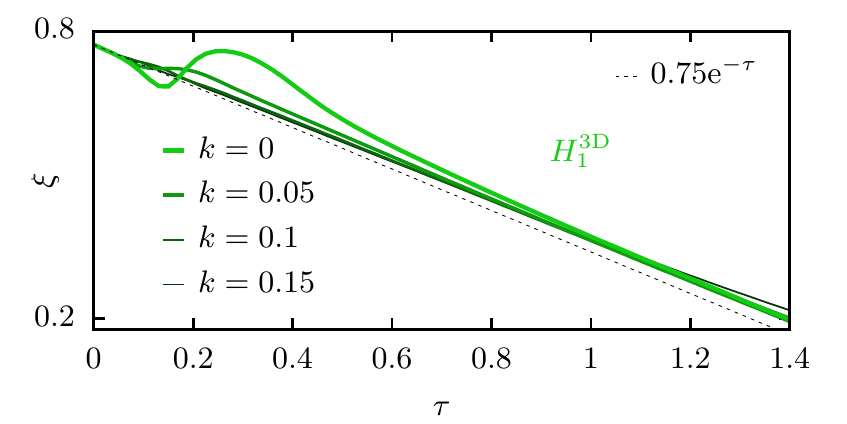}
\caption{\label{fig:3d_x_t} Propagation of wave packets in a 3D lattice with a horizon. The figure shows the center of mass of the wave packet along the $x$ axis as a function of time for different starting momenta in the systems $H_1^{\text{2D}}$ and $H_2^{\text{2D}}$ in Eq.~\eqref{eq:H2D}. The rescaled position and time are defined as $\xi=1-2x/L$ and $\tau=2t/L$. The dashed line shows the geodesic calculated from Eq.~\eqref{eq:geodesics}. The length, width and height of the system are $L=400$ (we only simulated the left half of the system), $W=300$ and $H=300$.}
\end{figure}

The world lines of the wave packets in 2D and 3D are shown in Figs.~\ref{fig:2d_x_t} and \ref{fig:3d_x_t}. Here we only focus on the center of mass of the wave packets on the $x$ axis. As we can see the results are very similar to the 1D results in Fig.~\ref{fig:1d_x_t}. The main difference are the oscillations due to the Zitterbewegung at small times.

\subsection{Power law horizons in 1D}

So far the tilting was always linearly dependent on the position, now we consider more generic cases and study how the different choices affect the wave packet propagation. We only discuss the $H_1^{\text{1D}}$ system, for the $H_2^{\text{1D}}$ model similar results can be obtained.

\begin{figure}
\centering
\includegraphics[width=8.6cm]{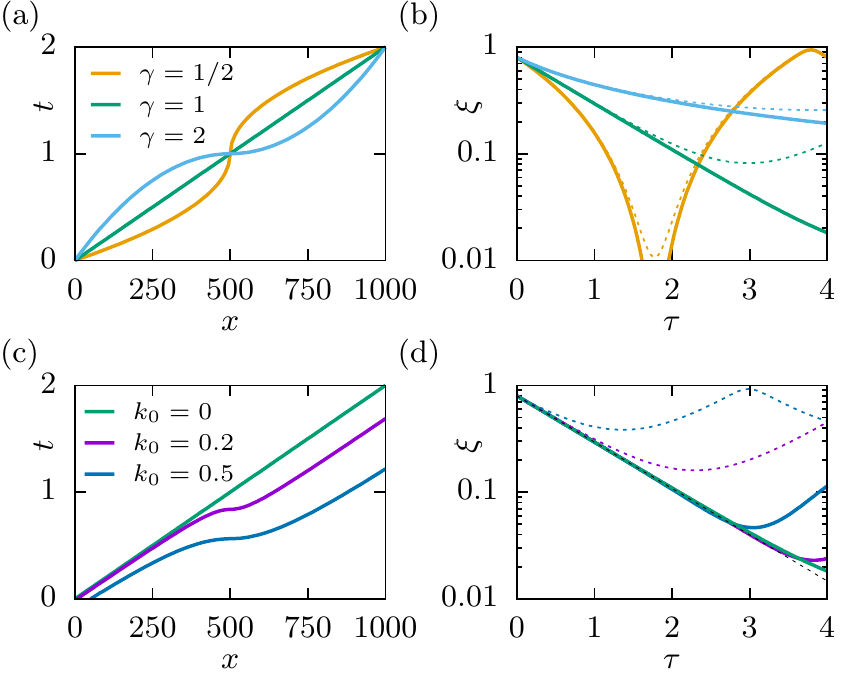}
\caption{\label{fig:1d_x_t_gamma} Propagation of wave packets in a 1D lattice with different types of horizons using the $H_1^{\text{1D}}$ system in Eq.~\eqref{eq:H1H2}. Panels (a,b) show the results for the tilting profile with different power laws as in Eq.~\eqref{eq:tgamma} for three different $\gamma$ exponents. Panels (c,d) show the results for the artificial horizons at finite momenta as in Eq.~\eqref{eq:tartificial} for three different $k_0$ values. Panels (a) and (c) show the position dependence of the tilting parameter in the different horizon models. Panels (b) and (d) show the center of mass of the wave packet as a function of time for the models corresponding to panels (a) and (c). In panel (b) solid lines are with $k=0$ and dashed lines with $k=0.1$. In panel (d) solid lines are at tilting profiles and momenta corresponding to panel (c), while dashed lines are at momenta corresponding to panel (c) but with the linear tilting profile. The rescaled position and time are defined as $\xi=1-2x/L$ and $\tau=2t/L$. The length of the system in all panels is $L=1000$.}
\end{figure}

We consider models that follow a power law close to the horizon where $t=1$. For the position dependence of the tilting parameter we use
\begin{equation}
\label{eq:tgamma}
    t(x) = 1+\mathrm{sgn}\left(2x-L\right)\left|\frac{2x}{L}-1\right|^\gamma ,
\end{equation}
where $\gamma$ is the exponent. At $\gamma=1$ we recover the model studied in the previous sections. This tilting dependence is shown in Fig.~\ref{fig:1d_x_t_gamma}(a). 

The world lines of the wave packet for different $\gamma$ parameters are shown in Fig.~\ref{fig:1d_x_t_gamma}(b). For linear and supralinear dependencies ($\gamma\geq1$) we get wave packets that infinitely approach the horizon. The higher the exponent the slower the approach is. For sublinear dependencies ($\gamma<1$) the wave packet goes very close to the horizon but ultimately bounces back from it. These results are consistent with the results obtained for the single band model in Ref.~\cite{Morice2021}.

\subsection{Horizons for finite energy in 1D}
In Sections \ref{sec:1dlin} we saw that in the $H_1^{\text{1D}}$ model, for the linear profile a wave packet at finite energies always deviates from the exponential time dependence and bounces back from the horizon. This happens because the effective horizon at finite energies is no longer at $t=1$ but at lower values, thus the wave packet can not reach the center of the chain. Focusing on the red band in Fig.~\ref{fig:disprel_1d}(a) the dispersion relation and group velocity are
\begin{subequations}
\begin{align}
    E &= (1-t)\sin{k},\\
    v&=(1-t)\cos{k},
\end{align}
\end{subequations}
which implies
$v^2={(1-t)^2-E^2}$.
Hence, for a wave packet of energy $E$ this means that there is an effective horizon at
\begin{equation}
    1-t(x_{\rm eff}) = \pm E.
\end{equation}

Knowing this we can construct a tilting profile that has a horizon at finite energy similar to the zero energy one. In order to reproduce the zero energy world line we want a linearly decreasing group velocity
\begin{equation}
    v(x) = 1 - \frac{2x}{L}.
\end{equation}
For $2x\leq L$ this can be satisfied by choosing
\begin{equation}
\label{eq:tartificial}
    t(x) = 1-\sqrt{\left(\frac{2x}{L}-1\right)^2+\left(\frac{2x_0}{L}-1\right)^2\tan^2{k_0}},
\end{equation}
where $k_0$ is the initial momentum and $x_0$ is the initial position of the wave packet. For $2x>L$ we can choose an arbitrary tilting profile since that part of the chain will not be accessible by the wave packet. For simplicity we use the same dependency but flipped. The tilting parameter as a function of position is shown in Fig.~\ref{fig:1d_x_t_gamma}(c). Using these horizons the wave packet world lines are shown in Fig.~\ref{fig:1d_x_t_gamma}(d). As we can see the world lines that were bouncing back with the linear profile at finite momenta now are approaching the horizon similarly to the $k=0$ wave packet. Close to the horizon because of lattice effects we get the disintegration of the wave packet similarly to the $k=0$ case.

\section{Scattering matrix}
\label{sec:S}
To characterize the different possible scenarios in the models discussed above we turn to scattering theory. Taking the same systems studied in the previous section we attach two semi-infinite leads on both ends. We restrict the discussion to 1D systems, because with the change in tilting always being in one direction only, we can always choose periodic boundary conditions in the other directions to get effectively 1D systems. The leads are made using the same model as the system with fixed tilting $t=0$ ($t=2$) for the left (right) lead. We then calculate the scattering matrix between the two leads numerically using the Kwant package \cite{Groth2014}. The scattering matrix ($S$) encodes the probability amplitudes of scattering from incoming modes ($\vb{i}$) into outgoing modes ($\vb{o}$).
\begin{equation}
    o_m = \sum\limits_n S_{mn}(\varepsilon)i_n,
\end{equation}
where $\varepsilon$ is the energy of the considered modes.

\begin{figure}
\centering
\includegraphics[width=8.6cm]{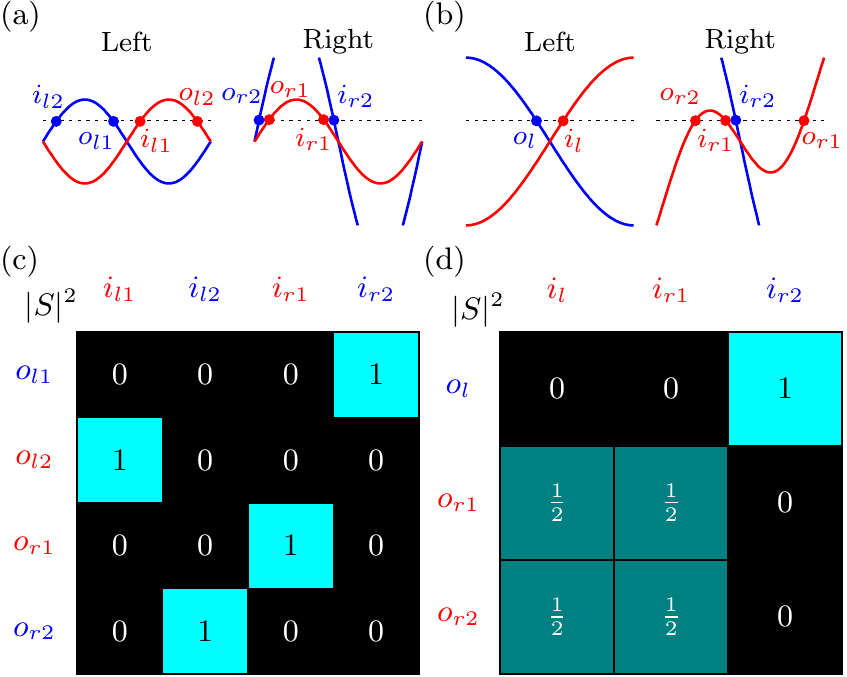}
\caption{\label{fig:1d_S} Propagating modes and scattering matrix of the $H_1^{\text{1D}}$ (a,c) and $H_2^{\text{1D}}$ (b,d) lattice models with a horizon. In panels (a) and (b) the dispersion relations in the left and right leads are shown. At the energy indicated by the dashed line the possible propagating modes are indicated by the filled circles. In panels (c) and (d) the scattering probabilities between the modes at zero energy are shown.}
\end{figure}

First, we discuss the $H_1^{\text{1D}}$ system. The propagating modes at a given $\varepsilon>0$ energy in this system are shown in Fig.~\ref{fig:1d_S}a. There are two incoming and two outgoing modes in each side of the scattering region resulting in a scattering matrix that is $4\times4$. The scattering matrix at $\varepsilon=0$ is shown in Fig.~\ref{fig:1d_S}c. We can see that each incoming mode is scattered into an outgoing mode with probability one. Incoming modes from the red bands are totally reflected into outgoing modes in the red band of the same lead. Incoming modes of the blue bands are completely transmitted into outgoing modes in the blue band of the opposite lead. This behavior is consistent with the reasoning of the previous section: the red band becomes completely flat so states in this band cannot cross the horizon, while the blue band stays qualitatively the same and the states in this band can go through the horizon. The same is valid for all energies as long as there are propagating modes in both leads.

Then, we study the $H_2^{\text{1D}}$ system. The propagating modes at a given $\varepsilon>0$ energy in this system are shown in Fig.~\ref{fig:1d_S}b. The left lead has one incoming and one outgoing mode while the right lead has two and two resulting in a scattering matrix that is $3\times3$. The scattering matrix at $\varepsilon=0$ is shown in Fig.~\ref{fig:1d_S}d. The incoming mode in the blue band is totally transmitted into the outgoing mode in the blue band. At zero energy the incoming modes in the red band are equally split between the two outgoing modes in the red band. In this system there is a big difference between the red incoming mode on the left and right sides. Starting from the left side we get total transmission, while starting from the right side we get total reflection.

\begin{figure}
\centering
\includegraphics[width=8.6cm]{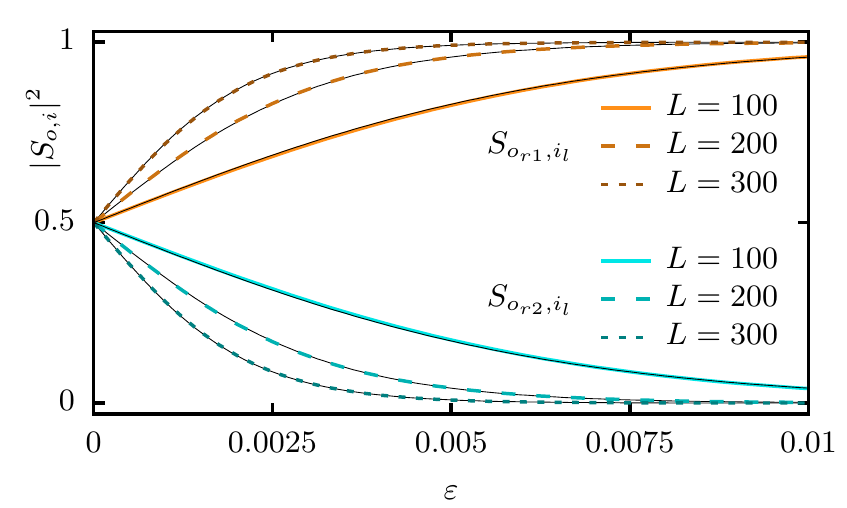}
\caption{\label{fig:1d_S_m1} Energy dependence of the scattering probability from the incoming mode $i_l$ to the outgoing modes $o_{r1}$ and $o_{r2}$ from Fig.~\ref{fig:1d_S}b. The colored lines show the numerical results for different system lengths. The solid black lines show the analytic formulas in Eq.~\eqref{eq:Soril}.}
\end{figure}

In the $H_2^{\text{1D}}$ system the energy dependence of the scattering amplitudes is more complicated than the $H_1^{\text{1D}}$ system. The zeros and ones are unaffected but the equal splitting between the red bands is only valid at $\varepsilon=0$. Let us consider the $i_l$ incoming mode. At a large enough positive energy if we follow the same energy modes throughout the scattering region we see that they appear at continuously increasing momenta until we reach a large enough tilting at the other side of the horizon where a second same energy mode appears at a very different negative momentum. Since large momentum changes are not allowed because of the slow varying of the tilting the $i_l$ mode will be totally transmitted into the $o_{r1}$ mode at large enough positive energies. The closer we go to zero energy we see that at the point of the appearance of the second mode with the same energy, the difference between the two momenta is decreased, and at $\varepsilon=0$ it is equal to zero. This means that the $i_l$ mode can scatter into both the $o_{r1}$ and $o_{r2}$ with continuous momentum change, thus they have equal scattering probabilities. In between the zero and the high enough energies the scattering probabilities show an exponential behavior (see Fig.~\ref{fig:1d_S_m1})

The same was observed for the $H_2^{\text{1D}}$ Hamiltonian for a similar tilting dependency in Ref.~\cite{Sabsovich2021,de2021artificial}. The energy dependence of the scattering rate can be expressed as
\begin{subequations}
\label{eq:Soril}
\begin{align}
    \left|S_{o_{r1},i_l}\right|^2 &= \frac{1}{1+\mathrm{e}^{\pi L \varepsilon}},\\
    \left|S_{o_{r2},i_l}\right|^2 &= 1-\frac{1}{1+\mathrm{e}^{\pi L \varepsilon}}.
\end{align}
\end{subequations}
We see that the momentum jumps are exponentially suppressed. Increasing the system size makes the change of tilting smoother, which in turn makes the quasi conservation of momentum stricter, and causes the probability amplitudes to reach their asymptotic behavior faster.
The {\it Hawking fragmentation} coined in Ref.~\cite{Sabsovich2021} can thus be understood as intermode scattering between multiple states at the same energy with different momenta. Increasing the energy of the incoming mode increases the momentum difference between the scattered modes, which leads to the energy dependence shown in Fig.~\ref{fig:1d_S_m1}.

\section{Discussion and Conclusions}

We introduced lattice models for tilted WSMs where the tilting varies smoothly with position. These models at small momenta effectively correspond to the massless Dirac equation in a curved spacetime background. To understand the role of high momenta and inter-Weyl-node effects, we studied two different types of systems: the first type ($H_1$) has $2^d$ nodes where $d$ is the dimension of the system, while the second type ($H_2$) only has a single Weyl node.

Using these models, we simulated wave packet propagations for three types of tilting profiles in 1D, 2D and 3D, using the Chebyshev expansion method.
When the tilting varies linearly with position, and in the zero-energy limit of both $H_1$ and $H_2$, the wave packets follow the geodesics of (1+1)D de Sitter spacetime, as found previously in a single-band model \cite{morice2021prr}.
At finite energies ($E>0$) though, $H_1$ and $H_2$ give very different behaviours. $H_1$ causes wave packets to bounce back, similarly to the single-band model, while $H_2$ causes wave packets to be transmitted.
This is explained by the differences in the band structure at large momenta, which allow for scattering to same-energy states in $H_2$ but not in $H_1$. When the tilting varies as a power law with position, we found that the transition between models with and without a horizon found in the single-band model is also present in two-band models. Finally, we designed a specific tilting profile that moves the horizon away from zero energy, and can be tuned to obtain eternal slowing down for any initial wave packet.

We showed that these results can be understood using the local dispersion relations, the conservation of energy, and only allowing continuous change for the quasimomentum.
To further show the validity of this approach, we computed the scattering matrix for 1D systems, which corresponds well to this analysis.
$H_1$ only has fully transmitted or reflected modes, and no scattering takes place.
While in $H_2$, due to the presence of multiple states at a single energy, intermode scattering does occur, in agreement with scattering amplitudes obtained in a related model \cite{Sabsovich2021}, which we now relate directly to wave packet trajectories.

In summary, our results bridge the gap between the single-band description and concrete proposals for the experimental realisation of Weyl gravitational analogues.
Indeed, they point to the rich physics arising from the presence of a second band, and the presence or absence of other Weyl nodes at large momenta.
These can in some cases dominate the observed phenomena and therefore indicate limitations of the analogy between tilted WSMs and gravitational systems which should be taken into account in experimental setups.
Finally, we identified the possibility of tuning the tilting profile to provide horizons for finite-energy wave packets, thus expanding the possibilities offered by this type of gravitational analogue.

\begin{acknowledgments}
The authors are grateful to I. C. Fulga for helpful discussions on the subject. We thank Ulrike Nitzsche for technical assistance.
\end{acknowledgments}

\appendix

\section{Connection between tight-binding and continuum models}
\label{app:discrete}
In this section we show how the tight-binding Hamiltonians in Eq.~\eqref{eq:realham} become the Eq.~\eqref{eq:hamiltonain_le} continuum model when going to vanishing lattice constants. We only show $H_1^{\text{1D}}$ in detail, $H_2^{\text{1D}}$ and higher dimensions can be derived similarly. The single particle tight-binding eigenvalue problem can be written as
\begin{equation}
    E\psi_n = (\sigma_x-t_n\sigma_0)\frac{\psi_{n+1}}{2i}- (\sigma_x-t_{n-1}\sigma_0)\frac{\psi_{n-1}}{2i}.
\end{equation}
We can regroup terms in the same equation as
\begin{equation}
    \frac{E}{a}\psi_n = (\sigma_x-t_n\sigma_0)\frac{\psi_{n+1}-\psi_{n-1}}{2ia}- \frac{t_n-t_{n-1}}{2ia} \psi_{n-1},
\end{equation}
where we divided the equation with the lattice constant $a$. Approximating derivatives with finite differences as
\begin{align}
    \partial_x \psi &\leftrightarrow \frac{\psi_{n+1}-\psi_{n-1}}{2a},\\
    \partial_x t &\leftrightarrow \frac{t_{n}-t_{n-1}}{a},
\end{align}
and rescaling the energy with the lattice constant we get the Hamiltonian in Eq.~\eqref{eq:hamiltonain_le}
\begin{equation}
    E\psi(x) = -i(\sigma_i-t_i\sigma_0)\partial_x\psi(x)-\frac{\sigma_0}{2i} \left(\partial_x t(x)\right)\psi(x).
\end{equation}
\section{Chebyshev expansion method}
\label{app:kpm}
The time evolution of wave packets was performed using the Chebyshev expansion method \cite{Kosloff1994,Weisse2006}. The method is based on the following expansion of the exponential function
\begin{equation}
\label{eq:kpm}
        \mathrm{e}^{-itx} = J_0(t)+2\sum\limits_{m=1}^\infty (-i)^mJ_m(t)T_m(x),
\end{equation}
where $J_m$ are the Bessel functions of the first kind and $T_m$ are the Chebyshev polynomials of the first kind. This formula is valid in the range $x\in[-1,1]$. The Chebyshev polynomials can be obtained using the following recurrence relation
\begin{subequations}
\label{eq:recur}
\begin{align}
        T_{m+1}(x)+T_{m-1}(x)&=2xT_m(x), \\
        T_0(x)&=1,\\
        T_1(x)&=x.
\end{align}
\end{subequations}

This expansion can be used to calculate the time evolution of $\ket{\psi(0)}$
\begin{equation}
        \ket{\psi(t)}=\mathrm{e}^{-itH} \ket{\psi(0)}.
\end{equation}
Using Eq. \eqref{eq:kpm} we get
\begin{equation}
        \ket{\psi(t)} = J_0(t)\ket{\psi(0)}+2\sum\limits_{m=1}^\infty (-i)^mJ_m(t)T_m(H)\ket{\psi(0)}.
\end{equation}
In order for the Chebyshev expansion to be convergent the Hamiltonian has to be normalized such that each eigenvalue is in the range $[-1,1]$. Using the recurrence relation in Eq.~\eqref{eq:recur}
\begin{subequations}
\begin{align}
        T_0(H)\ket{\psi}&=\ket{\psi},\\
        T_1(H)\ket{\psi}&=H\ket{\psi}, \\
        T_{m+1}(H)\ket{\psi}&=\left[2HT_m(H)-T_{m-1}(H)\right]\ket{\psi}.
\end{align}
\end{subequations}

This algorithm does not require the eigenvalue problem of the Hamiltonian to be solved. Using sparse matrices the recurrence relation can be evaluated efficiently, allowing us to study much larger systems than those accessible with methods that require the diagonalization of the Hamiltonian.

\bibliography{references}

\end{document}